\documentclass[nohyperref]{article}

\usepackage{microtype}
\usepackage{graphicx}
\usepackage{subfigure}
\usepackage{booktabs} 
\usepackage{longtable}
\usepackage{verbatim}
\usepackage{svg}
\usepackage{makecell}

\usepackage{pifont}

\usepackage{hyperref}

\usepackage[accepted]{icml2022}

\usepackage{amsmath}
\usepackage{amssymb}
\usepackage{mathtools}
\usepackage{amsthm}
\usepackage{multirow}

\usepackage[capitalize,noabbrev]{cleveref}

\theoremstyle{plain}

\theoremstyle{definition}

\theoremstyle{remark}

\usepackage{enumitem}

\usepackage[textsize=tiny]{todonotes}

\begin{document}

\twocolumn[
\icmltitle{Deduplicating Training Data Mitigates Privacy Risks in Language Models}








\begin{icmlauthorlist}
\icmlauthor{Nikhil Kandpal}{unc}
\icmlauthor{Eric Wallace}{berkeley}
\icmlauthor{Colin Raffel}{unc}
\end{icmlauthorlist}

\icmlaffiliation{unc}{UNC Chapel Hill}
\icmlaffiliation{berkeley}{UC Berkeley}

\icmlcorrespondingauthor{Nikhil Kandpal}{nkandpa2@cs.unc.edu}

\icmlkeywords{Language Models, NLP, Memorization, Privacy}

\vskip 0.3in
]

\printAffiliationsAndNotice{}
\begin{abstract}
Past work has shown that large language models are susceptible to privacy attacks, where adversaries generate sequences from a trained model and detect which sequences are memorized from the training set. In this work, we show that the success of these attacks is largely due to duplication in commonly used web-scraped training sets.
We first show that the rate at which language models regenerate training sequences is \emph{superlinearly} related to a sequence's count in the training set. 
For instance, a sequence that is present 10 times in the training data is on average generated ${\sim}1000\times$ more often than a sequence that is present only once.
We next show that existing methods for detecting memorized sequences have near-chance accuracy on non-duplicated training sequences.
Finally, we find that after applying methods to deduplicate training data, language models are considerably more secure against these types of privacy attacks.
Taken together, our results motivate an increased focus on deduplication in privacy-sensitive applications and a reevaluation of the practicality of existing privacy attacks.
\end{abstract}

\section{Introduction} \label{sec:intro}



\begin{figure}[t]
\centering
\includegraphics[trim={0.0cm 0cm 0.0cm 0.0cm},clip, width=1.0\columnwidth]{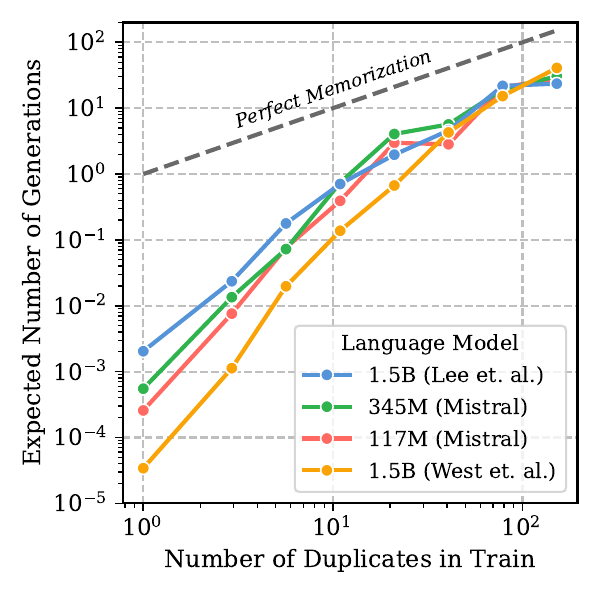}
\vspace{-1.0cm}
\caption{
For a sequence duplicated $d$ times in a language model's training dataset, we measure how often that sequence is expected to occur in a set of generated text that is equal in size to the training data. \emph{Perfect Memorization} amounts to generating a sequence at the same frequency as it appears in the training data. All LMs tested show a superlinear increase in the expected number of generations (slopes~$>1$ on a log-log plot), i.e., training samples that are not duplicated are very rarely generated, whereas samples that are duplicated multiple times appear dramatically more frequently.}
\label{fig:teaser}
\end{figure}

Neural language models (LMs)---systems trained to predict the next-word in a sequence of text---have become fundamental building blocks for numerous NLP tasks and domains. 
The performance and generality of these models make it important to study the extent to which they maintain the privacy of their training data, because many of their applications involve training on private information (e.g., emails, health records, chat logs, and source code).

Unfortunately, when training on private data, LMs may memorize and leak information to adversaries.  
Past work has demonstrated the practicality of these so-called model inversion attacks, which can successfully recover training data with only black-box access to a trained model \cite{carlini2019secret,carlini2021extracting,inan2021training}. In particular, the strongest attack, proposed by \citet{carlini2021extracting}, recovers training data from LMs by first generating sequences from the models and then scoring those sequences with various membership inference methods. The highest-scoring sequences are classified as belonging to the training data.

\begin{figure*}[htbp]
  \centering
  \includegraphics[trim={0.8cm 4.2cm 0.2cm 1.0cm},clip,page=4,width=0.9\textwidth]{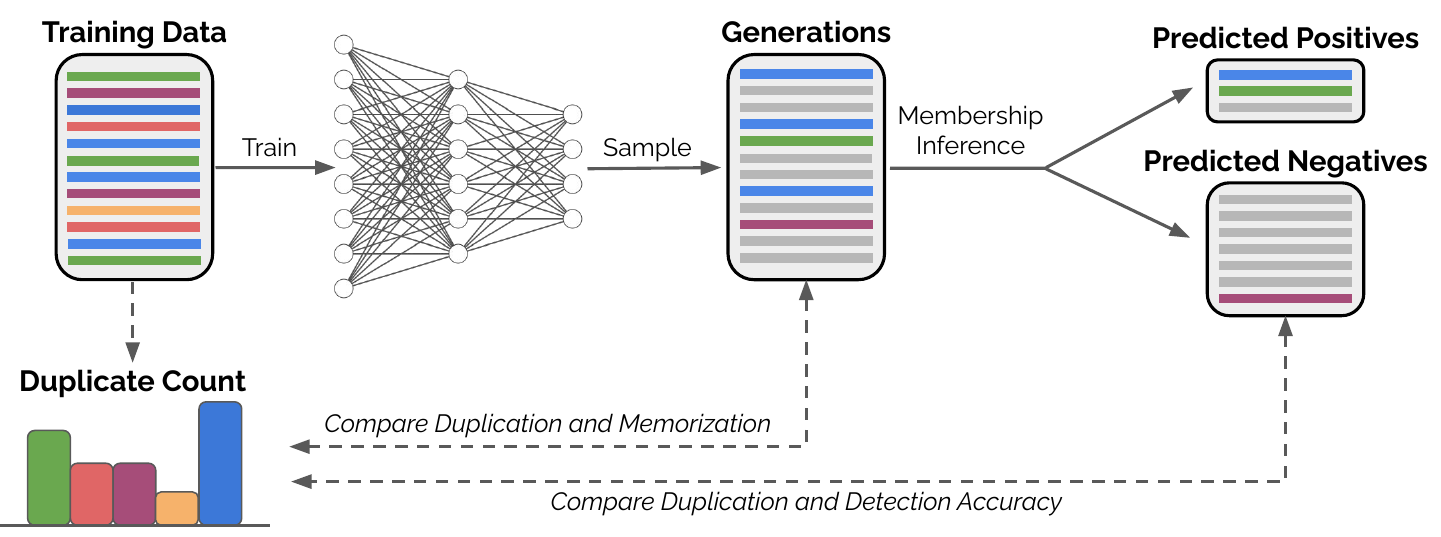}
  \caption{\textit{Overview of our analysis.} Web-scraped text datasets that are used to train language models contain duplicated sequences, depicted in the figure as training data rows of the same  color (\textit{top left}). Model inversion attacks attempt to recover training data from a trained model by first generating large amounts of text, some of which is memorized training data (\textit{top middle}). Membership inference is then performed to detect which generated sequences were copied from the training data (\textit{top right}). Our analysis focuses on the relationship between the amount a sequence is duplicated in the training data and the effectiveness of the model inversion attack at generating and detecting that sequence (\textit{bottom}).}
  \label{fig:block_diagram}
\end{figure*}


In this work, we show that the success of the \citet{carlini2021extracting} attack is largely due to duplicated sequences found in commonly used web-scraped training datasets. We study transformer LMs over various parameter scales and show that (1) the attack's likelihood of recovering a particular training sequence is correlated with the number of occurrences of that sequence in the training data, and (2) the overall attack effectiveness is reduced when sequence-level duplication in the training data is removed.

Concretely, we first show that the content that an LM generates is highly sensitive to sequence-level duplication in the training data. Using various sampling strategies, we generate text from LMs ranging from 117M-1.5B parameters. We consistently find a superlinear relationship between the number of times a sequence is duplicated in the training data and the rate at which that sequence is generated (e.g., Figure~\ref{fig:teaser}). For instance, a sequence that is present 10 times in the training data is on average generated ${\sim}1000\times$ more often than a sequence that is present only once.  Notably, our results show that \textit{samples which are not duplicated are very rarely regenerated by language models}.

We then look at the next stage of the model inversion attack: detecting memorized training data from a set of LM generations.
We demonstrate that the membership inference methods from \citet{carlini2021extracting} are correlated with the number of duplicates of a sequence in the training data. For example, the membership inference methods have an area under the ROC curve as high as 0.90 for sequences that are duplicated many times but \textit{achieve only chance accuracy for sequences that appear once}. 

In our final set of experiments, we directly test whether retraining LMs on deduplicated training datasets can mitigate privacy risks. We find that model inversion attacks are indeed much weaker for deduplicated models: \textit{they emit ${\sim}20\times$ less training data and reduce the effectiveness of two of the three proposed membership inference methods}. All in all, our results underscore the need to carefully remove duplicates when training privacy-sensitive models and show that past work may overestimate the effectiveness of LM privacy attacks when duplication is mitigated. 
\section{Background and Experimental Setup}

\paragraph{Language Models} Language models take as input a sequence of tokens and output a probability distribution over the next token. 
LMs are trained to maximize the likelihood of a corpus of text and can be used to generate text at test time by iteratively sampling from the next-token distribution. In practice, various strategies exist for sampling tokens, including random sampling, sampling from the top-$k$ highest probability tokens~\cite{fan2018hierarchical}, or sampling after using a temperature to sharpen the next-token distribution.

\paragraph{Memorization}
The concept of ``memorization'' refers to ways that a trained model stores and consequently leaks information about its training data. Multiple notions of memorization have been studied that vary in their definitions and assumptions (see Section~\ref{sec:related_work} for further discussion). In this work we focus on generation-based memorization, where a generative model leaks information by generating exact samples from its training data~\cite{carlini2019secret}.

When studying generation-based memorization in LMs, we compare models' generation behavior with the expected behavior of a model that has perfectly fit the training data through memorization. This \textit{perfect memorization} model only assigns non-zero probability to samples seen during training and sampling from the model is identical to uniformly sampling from the training data. The perfect memorization model serves as a positive control showing how far LMs are from fully memorizing their training data.

\begin{figure*}[htp]
\centering
\subfigure[]{
    \includegraphics[trim={0.1cm 0cm 1.1cm 0.6cm},clip, width=0.48\textwidth]{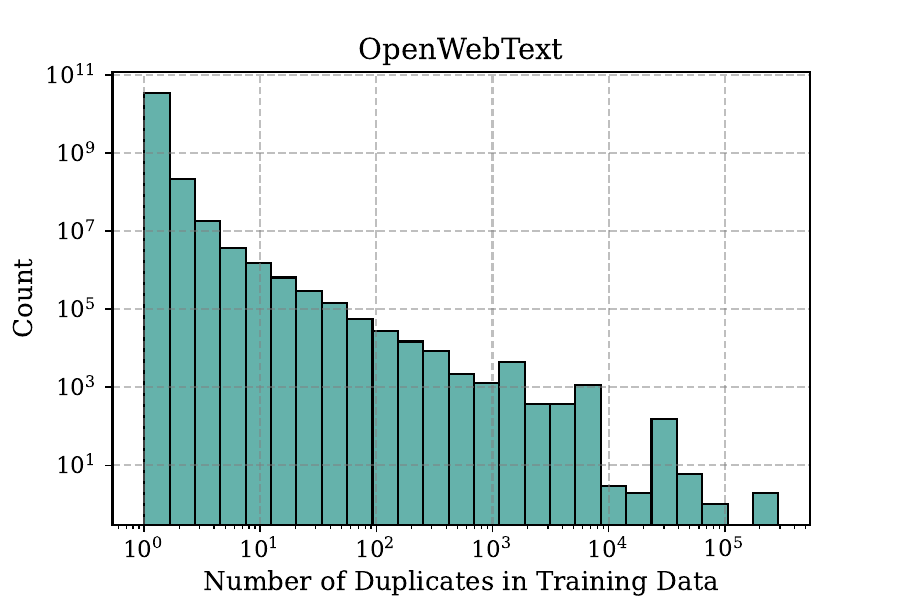}
    \label{fig:owt_histogram}
    }
\subfigure[]{
\includegraphics[trim={0.1cm 0cm 1.1cm 0.6cm},clip,width=0.48\textwidth]{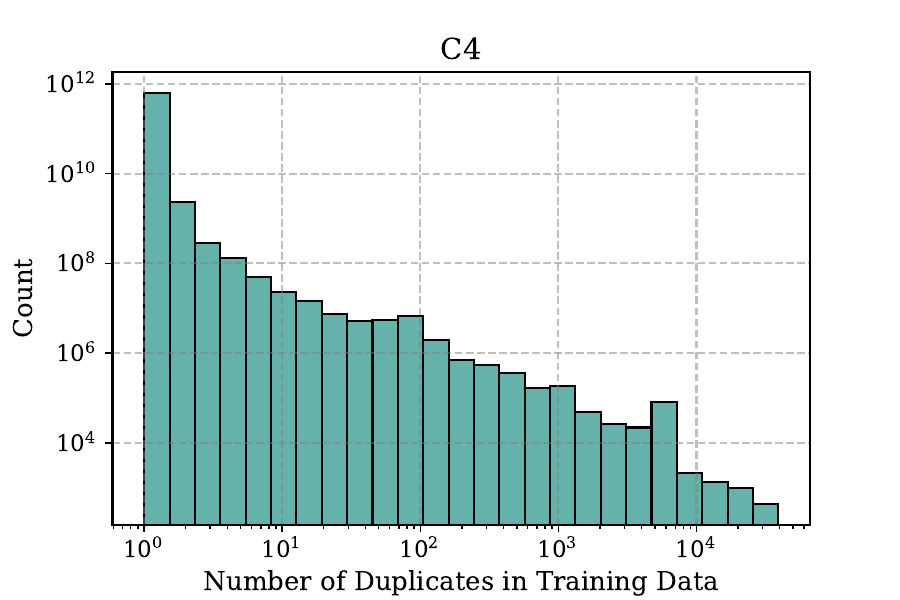}
  \label{fig:c4_histogram}
} 
\vspace{-0.45cm}
\caption{Web-scraped training sets are rife with duplicated sequences. Above, we plot the frequency of different amounts of duplication for 400-character sequences in the OpenWebText \textbf{\subref{fig:owt_histogram}} and C4 \textbf{\subref{fig:c4_histogram}} datasets. Note that C4 is an order of magnitude larger than OpenWebText.}
\label{fig:histograms}
\end{figure*}

\paragraph{Privacy Attacks} The fact that state-of-the-art LMs memorize and regenerate sequences seen during training enables attacks that compromise the privacy of their training data~\cite{carlini2019secret,inan2021training,carlini2021extracting}. In this work, we focus specifically on the \citet{carlini2021extracting} attack, which is currently the strongest and most accessible model inversion attack on LMs. While we focus on this particular attack, our analysis also applies to other attacks that leverage generation-based memorization.

The \citet{carlini2021extracting} attack works in two stages:
\begin{enumerate}[leftmargin=15pt,itemsep=0mm]
    \itemsep3pt 
    \item Generate a large amount of text from a language model.
    \item Score the generated sequences using a membership inference scoring method.
\end{enumerate}

For the first stage, \citet{carlini2021extracting} study different methods of generating data (unconditional vs. conditional sampling, different sampling strategies, etc.). We focus on unconditional generation using standard sampling, top-$k$ sampling, and temperature sampling. 

For the second stage, we study all scores proposed by \citet{carlini2021extracting}. Each score is defined as the ratio between a metric estimating the ``easiness'' of sequence (a property of the sequence itself and not whether the sequence appears in the training dataset) and the trained model's perplexity on that sequence. For measures of easiness, \citet{carlini2021extracting} used three choices:
\vspace{-0.3cm}
\begin{itemize}[leftmargin=15pt,itemsep=0mm]
    \item {\fontfamily{lmtt}\selectfont \textbf{Reference Model}}: the perplexity of another LM on the sequence. We use the GPT-2 small language model~\cite{Radford2019LanguageMA}.
    \item {\fontfamily{lmtt}\selectfont \textbf{zlib}}: the length of the sequence after compression by the zlib compression library.
    \item {\fontfamily{lmtt}\selectfont \textbf{Lowercase}}: the trained model's perplexity on the sequence with all lowercased characters.
\end{itemize}

\paragraph{Training Data Collection and Duplication} 
Modern language modeling datasets are generated by large-scale scraping of the Internet~\cite{Gokaslan2019OpenWeb,Radford2019LanguageMA,raffel2020exploring,gao2020pile}. Most web-scraped datasets are deduplicated at the \textit{document level}, e.g., if two web pages have the exact same contents, only one is kept in the data. \citet{lee2021deduplicating} observe that these datasets still have large-scale approximate and exact sequence-level duplication, e.g., quotes, paragraphs, and advertisements appearing in many web pages. To correct this, they propose efficient sequence-level deduplication methods based on locality sensitive hashing and suffix arrays.

When measuring duplication in the training data, we consider identical sequences to be duplicates. Although broader definitions such as approximate or semantic duplication may also be useful to study, we choose to investigate exact duplication in this work because it matches the adversary's goal of \emph{exactly} recovering sequences from the training data (e.g., social security numbers).
To detect duplicate sequences, we adapt the suffix array-based algorithm from \citet{lee2021deduplicating}. Searching for exactly duplicated sequences in two sets of text can be done efficiently with a linear traversal of the two texts' suffix arrays.

\begin{figure}[t]
\centering
\includegraphics[trim={0.0cm 0cm 0.0cm 0.0cm},clip, width=1.0\columnwidth]{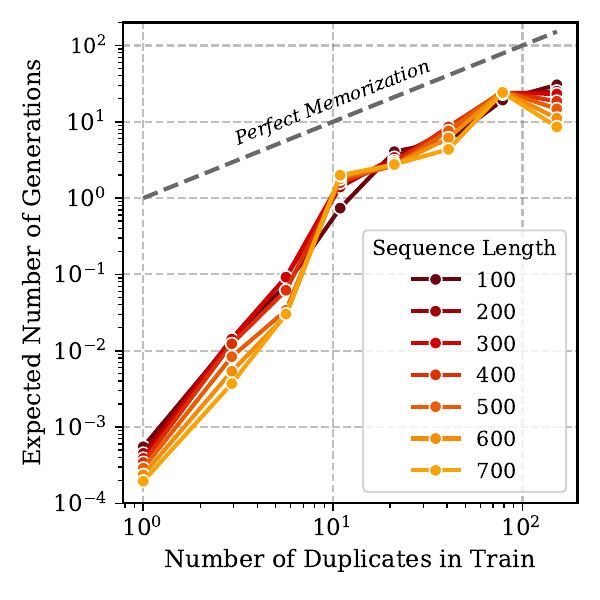}
\vspace{-0.85cm}
\caption{We vary the sequence length that is used when measuring whether a model generation overlaps with the training set. Using longer sequence lengths naturally reduces the chance that a generation exactly overlaps with the training set. However, the overall shape of the generation vs. duplication curve is consistent across a range of sequence lengths.}
\label{fig:seq_length}
\end{figure}

\paragraph{Datasets}
In our experiments, we use models trained on the widely-used OpenWebText~\cite{Gokaslan2019OpenWeb} and C4~\cite{raffel2020exploring} datasets. Both are large-scale datasets, 39GB and 750GB respectively, and were generated by scraping text from the Internet with basic filtering and deduplication. Despite deduplicating at the level of whole training examples, both datasets still contain a large number of duplicated token sequences between training examples, a property that is not unique to just these two datasets \cite{lee2021deduplicating}. To illustrate this quantitatively, in Figure~\ref{fig:histograms} we show how often each unique 400-character sequence is duplicated in these two datasets. Both datasets contain millions of sequences that are duplicated 10 or more times, and some individual sequences are even duplicated tens of thousands of times. This large amount of sequence-level duplication allows us to reliably measure the effect of duplication on memorization and downstream model privacy over a wide range of duplication levels.

\paragraph{Models} We focus on Transformer-based~\cite{vaswani2017attention} language models that range in scale from millions to billions of parameters. Specifically,
we use the 117M and 345M parameter models from the Mistral project\footnote{\url{https://github.com/stanford-crfm/mistral}} and the 1.5B parameter forward language model from \citet{west2021reflective}, all of which were trained on the OpenWebText dataset. Additionally, we evaluate the two 1.5B parameter models from \citet{lee2021deduplicating}, one trained on the C4 dataset and the other trained on a sequence-level deduplicated version of C4. We choose this set of models as they are near-state-of-the-art, and they allow us to test the effect of model scale, changes in codebase and implementation, optimization hyperparameters, and training data.

\begin{figure*}[htp]
\centering
\subfigure[]{
    \includegraphics[trim={0.2cm 0.5cm 0.1cm 0.35cm},clip, width=0.48\textwidth]{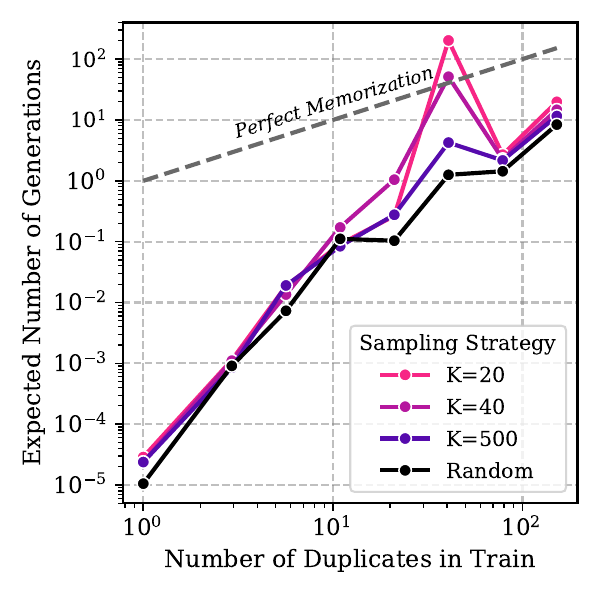}
    \label{fig:top_k}
    }
\subfigure[]{
\includegraphics[trim={0.2cm 0.5cm 0.1cm 0.2cm},clip,width=0.48\textwidth]{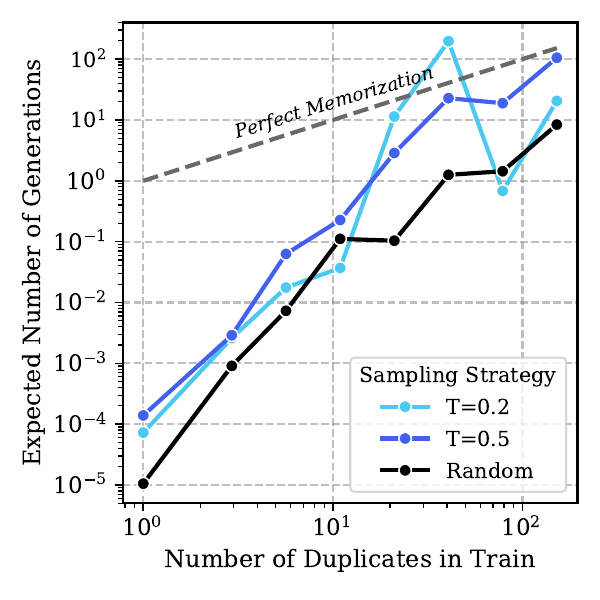}
  \label{fig:temp}
} 
\vspace{-0.35cm}
\caption{The sampling method impacts how often LMs regenerate training samples. Sampling methods that emit more likely sequences (e.g., top-$k$ with smaller $k$ or temperature sampling with smaller $T$) generate more verbatim training samples. Nevertheless, all sampling methods rarely generate training sequences when the number of duplicates is small.}
\label{fig:sampling_strategy}
\end{figure*}

\paragraph{Experimental Setup}
Our experiments follow the analysis depicted in Figure \ref{fig:block_diagram}. In Section \ref{sec:generation} we analyze the likelihood of regenerating a training sample as a function of that sample's number of duplicates in the training data. In Section \ref{sec:detection}, we analyze the relationship between duplication and the detection of LM generations copied from the training data. The code used to perform our experiments can be found at \url{https://github.com/nkandpa2/lm_memorization}.
\section{How Duplication Affects The Regeneration of Training Sequences} \label{sec:generation}
The first step of the \citet{carlini2021extracting} attack is to generate a large pool of sequences in hopes that some are verbatim copies from the training data. In this section, we analyze how duplication in the training data affects this stage.

Concretely, we first record the number of duplicates for each $N$-length character sequence in the training data. We then generate many times from an LM and analyze how often each $N$-length training sequence is generated as a function of its duplicate count.
Note that we also scale our calculations to simulate a scenario where we generate an amount of text equal in size to the training dataset. This allows us to directly compare the behavior of models trained on datasets of different sizes, and also compare to a theoretical model that has perfectly memorized its training data (i.e., generating from this model is identical to sampling from the training dataset). 

\subsection{Regeneration is Superlinearly Related to Duplicates}

All models that we test have a superlinear relationship between the number of times a training sequence is regenerated and the number of times that sequence is duplicated in the training data. This relationship is shown in Figure \ref{fig:teaser} by the $>1$ slope on a log-log plot. 

Furthermore, Figure \ref{fig:teaser} shows that the generation behavior of LMs is far from perfect memorization: sequences duplicated $d$ times in the training data are expected to be generated far fewer than $d$ times by a trained model. This is especially true for low duplicate counts, i.e., \textit{samples which are not duplicated are very rarely regenerated by language models}. This shows that the \citet{carlini2021extracting} attack---which relies on models regenerating training samples---will rarely be able to extract training data that is not duplicated.

Our finding that LMs exhibit a superlinear increase in their regeneration rates is a also novel phenomenon worthy of future study. Concretely, one would expect that LMs would exhibit ``calibrated'' generation behavior---training sequences that appear twice as frequently are twice as likely to be generated---but this is not true for state-of-the-art models.

\subsection{Regeneration Trends Are Robust Across Experimental Setups}

Having observed an initial superlinear trend, we next measure whether this relationship is a more general phenomenon that holds across different experimental setups varying the sequence length, model size, sampling strategy, and number of training epochs.

\paragraph{Effect of Duplicate Sequence Length} Our analysis focuses on the duplication of $N$-length character sequences in the training data. To ensure that our conclusions are not dependent on any one choice of $N$, we vary the sequence length and report the results for the Mistral 345M parameter model in Figure \ref{fig:seq_length}.  Using
longer sequence lengths naturally reduces the chance that a generation exactly overlaps with the training set. Nevertheless, the superlinear relationship between generation and duplication is nearly identical across different sequence lengths. For the rest of the paper we set $N = 100$ characters unless otherwise specified.

\paragraph{Effect of Model Scale} Larger models tend to regenerate more training data across all levels of duplication. This effect is shown by comparing the duplication curves for the 117M and 345M parameter Mistral models in Figure~\ref{fig:teaser}. These two models were trained nearly identically and thus the comparison controls for confounding factors such as the number of training steps and optimization hyperparameters. Larger models likely regenerate more training sequences because they achieve a lower training loss (i.e., they assign training samples higher likelihoods on average).

\paragraph{Effect of Sampling Scheme}
The amount of regeneration depends on the sampling scheme used. Figure \ref{fig:sampling_strategy} compares random, top-$k$, and temperature sampling for the Mistral 117M parameter model. We find that sampling schemes that emit more likely sequences (e.g., top-$k$ with smaller $k$) generate more verbatim training samples.

\begin{figure}[t]
\centering
\includegraphics[trim={0.0cm 0cm 0.0cm 0.0cm},clip, width=1.0\columnwidth]{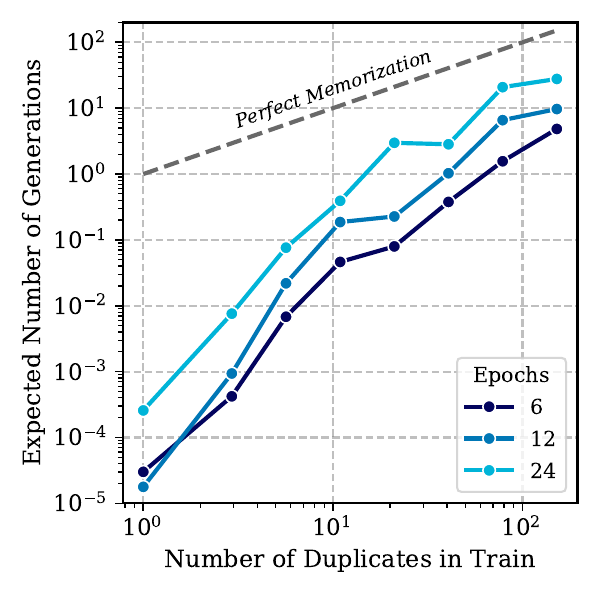}
\vspace{-0.8cm}
\caption{We plot the effect of performing multiple training epochs on the generation behavior. Performing additional epochs has a multiplicative effect that is uniform across all duplication levels. In particular, using twice as many epochs will cause the expected number of generations to increase by approximately 3 times for all duplication levels.}
\label{fig:epochs}
\end{figure}

\paragraph{Effect of Increasing Epochs}
Finally, we find that the regeneration rate increases over the course of training. Figure \ref{fig:epochs} shows that as training progresses for the 117M parameter Mistral model, the regeneration rate of training sequences increases at nearly all levels of duplication. Notably, stopping early does not change the fact that language models generate disproportionately many highly-duplicated training sequences.

\section{How Duplication Affects The Detection of Training Sequences}\label{sec:detection}

\begin{figure*}[htp]
\centering
\subfigure[]{
    \includegraphics[trim={0.2cm 0.5cm 0.1cm 0.35cm},clip, width=0.48\textwidth]{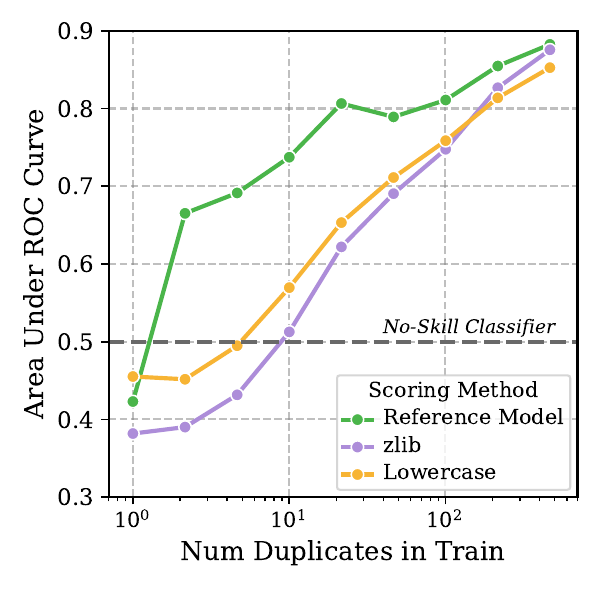}
    \label{fig:auroc}
    }
\subfigure[]{
\includegraphics[trim={0.2cm 0.5cm 0.1cm 0.2cm},clip,width=0.48\textwidth]{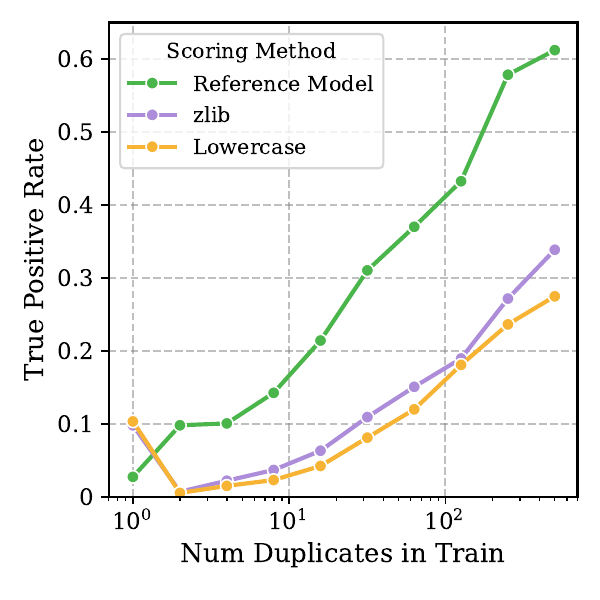}
  \label{fig:tpr}
} 
\vspace{-0.4cm}
\caption{State-of-the-art membership inference methods fail to accurately detect training sequences when they are not duplicated in the training set. In \textbf{\subref{fig:auroc}}, we report the area under the ROC curve for different membership inference methods as a function of the number of duplicates. In \textbf{\subref{fig:tpr}}, we report the true positive rate at a false positive rate of 0.1\%.}
\label{fig:intro}
\end{figure*}

Thus far, we found that models rarely regenerate training sequences that are not duplicated many times. Nevertheless, the second stage of the \citet{carlini2021extracting} attack, which looks to identify training sequences using membership inference methods, may be able to flag these rare cases. To test this, we evaluate the three membership inference scoring methods ({\fontfamily{lmtt}\selectfont Reference Model}, {\fontfamily{lmtt}\selectfont zlib}, and {\fontfamily{lmtt}\selectfont Lowercase}) and stratify the results by different duplication levels.

Concretely, we bucket the samples generated from the 345M parameter Mistral model into sequences that are duplicated in the training data $d$ times, for $d = 1$ to $d=800$. We also collect a set of 25,000 negative sequences that were generated by the LM but were not in the training data. Using these two sets of samples, we measure the effectiveness of the three membership inference scores at distinguishing between the two sets.

Figure \ref{fig:auroc} shows the area under the Receiver Operating Characteristic (AUROC) curve achieved by the different membership inference scores. Notably, for generated sequences only found once in the training data, all three scores yield classifiers that are close to chance.\footnote{A random ``no-skill classifier'' has an AUROC of 0.50.} Of the three scores, the {\fontfamily{lmtt}\selectfont Reference Model} is the highest performing classifier at nearly all levels of duplication.

Following the suggestions of \citet{carlini2021membership}, we also evaluate membership inference using the True Positive Rate (TPR) at a very low False Positive Rate (FPR). This simulates the realistic evaluation setting where the prevalence of training data in a set of generated samples is very low compared to non-training data. We found that approximately 1 in 1000 generated 100-character spans are copied from the training data. Thus, we use a FPR of 0.1\%. 

In Figure \ref{fig:tpr}, we show that the TPR of all three membership inference scores are highly correlated with the number of duplicates. For example, the TPR for the {\fontfamily{lmtt}\selectfont Reference Model} method is as high as 0.60 for sequences that are duplicated many times but is only 0.10 for sequences that appear once. All in all, these results show that while membership inference methods may achieve non-trivial accuracies on average over the entire generated set, \textit{most of their successes are on sequences that have been duplicated many times.}

\section{Model Inversion with Deduplicated Data}

In our final set of experiments, we directly test whether retraining LMs on deduplicated data can indeed mitigate privacy risks. In particular, we test two of the 1.5B parameter LMs from \citet{lee2021deduplicating}, one trained on C4 and another trained on a deduplicated version of C4.\footnote{We use the LM trained on the version of C4 that has exact duplicates removed using the suffix array-based \textsc{ExactSubstr} method. This removes exact duplicates that were at least 50 byte-pair encoding (BPE) tokens long. To ensure that we do not attempt to recover sequences shorter than 50 tokens, we set $N = 400$ characters in this experiment.}

\paragraph{Generating From Deduplicated Models} We first generate one million samples from each of the language models and measure the number of sequences copied from the training data. The top of Table~\ref{tab:model_inversion} shows the number of unique 400-character training sequences generated by each of the language models (\textit{Count}) and the percentage of all 400-character training sequences that are generated (\textit{Percent}). Respectively, these measure the total amount of information from the training data leaked by each model and the probability of a single sequence in the training data being leaked.
We find that the model trained on deduplicated data emits ${\sim}20\times$ less training data, i.e., deduplication strongly weakens the first stage of the \citet{carlini2021extracting} attack.

\paragraph{Membership Inference}
Next, we evaluate the performance of the membership inference scoring methods on the generated samples. We randomly subsample 25,000 sequences that are copied from the training data and 25,000 sequences that are novel from each of the models. All of these sequences are scored by the membership inference methods and we report the AUROC in the bottom of Table~\ref{tab:model_inversion}. 
We find that {\fontfamily{lmtt}\selectfont zlib} and {\fontfamily{lmtt}\selectfont Lowercase} are considerably affected by deduplication, whereas {\fontfamily{lmtt}\selectfont Reference Model} performs almost equally as well on both models. 

One factor to consider when comparing the AUROC scores between the normal and deduplicated models is that the set of memorized training samples that they are trying to detect are different. We hypothesize that in the rare circumstance that the deduplicated model does indeed regenerate a training sample, those samples may be unique in some manner. This may make the samples easier to classify, which can explain why the AUROC can remain relatively high for the deduplicated model (0.87). Further investigation of the difference between the regenerations made by a normal and deduplicated model is worthy of future study. 

\paragraph{Qualities of Effective Membership Inference Methods} The reasonable accuracy of the {\fontfamily{lmtt}\selectfont Reference Model} method suggests that it measures training data leakage beyond just generation-based memorization. We hypothesize that this is due to its similarity to a different notion of memorization known as counterfactual memorization \cite{zhang2021counterfactual}. A sample's counterfactual memorization is measured by comparing the sample's expected likelihood under models that have and have not trained on that sample. The {\fontfamily{lmtt}\selectfont Reference Model} method is an approximation of counterfactual memorization that uses a single model trained on different training data to approximate the expected likelihood under a model that has not trained on the sample being scored. While we find that duplication and generation-based memorization are highly correlated, this result suggests that approximating other notions of memorization, such as counterfactual memorization, may lead to membership inference scores that are less sensitive to deduplication. Similar findings have been noted in \citet{watson2021importance} and \citet{carlini2021membership}.

\paragraph{Is Deduplication An Effective Defense?}
Overall, our results show that deduplication is an effective safeguard against models \textit{regenerating} their training data, which renders the first stage of many existing model inversion attacks largely ineffective. Fortunately, this defense comes at little-to-no cost in model performance, as training on deduplicated data does not harm language modeling perplexity \cite{lee2021deduplicating}.
Nevertheless, in the rare cases when deduplicated models do generate training data, those samples can still be detected somewhat reliably by membership inference scores such as the {\fontfamily{lmtt}\selectfont Reference Model} method.

\begin{table}[t]
\small
\centering
\begin{tabular}{clcc}
 & & \thead{Normal \\ Model} & \thead{Deduped \\ Model} \\
\midrule
\multirow{2}{*}{\makecell{Training Data \\ Generated}}
& \multicolumn{1}{|l}{Count} & 1,427,212 & 68,090 \\
& \multicolumn{1}{|l}{Percent} & 0.14 & 0.007 \\
\midrule
\multirow{3}{*}{\makecell{Mem. Inference \\ AUROC}}
&  \multicolumn{1}{|l}{zlib} & 0.76 & 0.67 \\
&  \multicolumn{1}{|l}{Ref Model} & 0.88 & 0.87 \\
&  \multicolumn{1}{|l}{Lowercase} & 0.86 & 0.68 \\
\bottomrule
\end{tabular}
\caption{Deduplicating training data drastically reduces the effectiveness of privacy attacks. We first generate 1 million 256-token samples from models trained on C4 and deduplicated C4. We then report the number of unique 400-character training sequences that are generated (\emph{Count}) and the percentage of all 400-character training sequences that are generated (\emph{Percent}). We then report the classification AUROC achieved by each of the three membership inference scores when applied to the generated sequences.}
\label{tab:model_inversion}
\end{table}
\section{Discussion} \label{sec:discussion}

\paragraph{More General Notions of Duplication} We define duplicates as two sequences that \emph{exactly} match one another. We chose this definition because it mirrors an adversary's goal of exactly recovering a training sequence. However, privacy can also be compromised by approximately recovering a training sequence. To study this, one would need to analyze near-duplicates. This is a challenging open problem as it can be difficult to detect more general notions of duplication such as sequences with similar semantics but different lexical forms~\cite{cer2017semeval}.

\paragraph{Duplication and Differential Privacy} Satisfying a strong differential privacy (DP) guarantee is considered the gold standard of protecting privacy \cite{dwork2006differential}. DP guarantees that the effect of a single training sample on a model is small. However, even when training with a strong DP guarantee, data points that are exact or near-duplicates can still possibly have a large cumulative impact on the model. Consequently, deduplication is still necessary even when training with DP.

\paragraph{Duplication Beyond Text Data} Our work focuses on natural language, but datasets in domains such as images and source code also contain duplicate samples \cite{recht2018cifar10,ziegler2021rote}. Models trained on these datasets have been shown to be vulnerable to data privacy attacks. However, it remains unclear as to whether the success of these attacks is mainly due to training data duplication. Given the results of our work, it is important to evaluate the relationship between duplication and privacy in non-language domains.
 \section{Related Work}\label{sec:related_work}
 
\paragraph{Memorization of Training Data} \label{paragraph:memorization}
Our work is enabled by models ``memorizing''' their training data. We focus on a definition of memorization that is based on regeneration of training data. Past and concurrent work uses similar definitions and experimental setups~\cite{mccoy2021much,lee2021deduplicating,carlini2022quantifying}. \citet{mccoy2021much} observe that LMs are capable of regenerating sequences over 1,000 words long. \citet{lee2021deduplicating} find that models trained on sequence-level-deduplicated data regenerate approximately 10 times less training data. Concurrent work from \citet{carlini2022quantifying} measures the worst-case memorization of language models by conditioning on prefixes from the training data. They find that the likelihood of a model generating exact continuations from the training data scales with model size, training data duplicates, and prefix length. Compared to these results, our work studies how sequence-level duplication affects the performance of practical privacy attacks that leverage this type of memorization.

Past work has also proposed alternate definitions of memorization. \citet{feldman2020neural} and \citet{burg2021memorization} define \textit{counterfactual memorization} as the difference between a training example's expected loss under models that have and have not been trained on that example. \citet{zhang2021counterfactual} study this form of memorization in large LMs. They find that training examples that are the most memorized are qualitatively different from other examples in the training set but simple enough to learn from a single training example. For long-tailed data distributions, counterfactual memorization can be necessary for learning accurate models \cite{feldman2020neural,brown2021when}. Our work does not focus on this definition of memorization as measuring it requires access to the training corpus and thus does not elicit practical privacy attacks.

\paragraph{Privacy Attacks}
Training data privacy can be compromised through membership inference attacks \cite{shokri2017membership}, which use a trained model to identify training data from a candidate set of samples. Past works on membership inference find that while overfitting is sufficient for performing membership inference, well-generalized models can also leak membership information \cite{yeom2018privacy,Long2018UnderstandingMI}. Membership inference can also be extended to audit models subject to data-protection laws \cite{song2018auditing}.  

Another type of privacy attack is model inversion. Early model inversion attacks use a trained model and non-sensitive features of a training sample to reconstruct that sample's sensitive features \cite{fredrikson2015model}. Later model inversion attacks focus on fully recreating training samples given access to only a trained model \cite{hidano2017model,song2020information,yang2019adversarial}. Autoregressive and masked transformer LMs have both been shown to be susceptible to model inversion \cite{carlini2021extracting,lehman2021clinical}. We build on \citet{carlini2021extracting}, who propose a model inversion attack that first generates a set of candidate samples from an autoregressive LM and then scores the generations based on their likelihoods relative to a baseline model.

\paragraph{Privacy Defenses}
Training data privacy can be protected using the differential privacy (DP) framework \cite{dwork2006differential}, which guarantees that the effect of any single training example on the trained model is not too large. \citet{yu2021differentially,li2021large} demonstrate the practicality of training differentially private LMs. \cite{zhao2022provably} propose provable confidentiality, a related guarantee that ensures that the content of particular secrets in the training data do not have a large effect on training. Other approaches such as \citet{mireshghallah2021privacy,li2018robust,coavoux2018privacy} use adversarial training to make private information more difficult to recover from model activations.

\paragraph{Benefits and Drawbacks of Deduplication}
\citet{lee2021deduplicating} study the effects of performing sequence-level deduplication on training corpora. They find that deduplication reduces the amount of training data emitted by trained LMs and speeds up the training process without harming model perplexity. \citet{hernandez2022scaling} also show that LM perplexity is harmed by data duplication, but only for an intermediate amount of duplication. They conjecture that this occurs when the amount of duplicated data is small enough to be memorized but large enough to use a significant amount of the model's capacity. Deduplication between a model's train and test set is also necessary for proper evaluation \cite{lee2021deduplicating,brown2020language}. 

Prior work using LMs for closed-book question answering shows that deduplication is not universally beneficial, as memorization of facts from the training data can be necessary for certain tasks \cite{petroni2019knowledge,roberts2020knowledge}.
\section{Conclusion and Future Work}
To create privacy-preserving machine learning models, one must go beyond simply identifying privacy vulnerabilities and instead trace the causes of vulnerabilities back to the training algorithms, models, and datasets. We take a step towards this goal by highlighting that sequence-level duplication is a large factor behind the success of recently proposed privacy attacks on LMs.
Moreover, our finding that LMs exhibit a superlinear increase in their regeneration rates as the number of duplicates increase is a novel phenomenon worthy of future study. 

We also show that past work may overestimate the effectiveness of privacy attacks when duplicates are removed from the training data. Consequently, future attack evaluations should take into account duplication as a possible confounding factor. More broadly, future attacks should be evaluated as a function of different features of the data, be it duplication or otherwise. This will allow a better understanding of when attacks succeed and how to defend against them.
\section*{Acknowledgements}
We thank Katherine Lee, Daphne Ippolito, Nicholas Carlini, and Adam Roberts for giving feedback on our work and providing access to the language models trained on C4 and deduplicated C4.
\newpage

\bibliography{paper}
\bibliographystyle{icml2022}


\end{document}